\documentclass[onecolumn,a4paper,fleqn,usenatbib]{mnras}
\usepackage{amsmath}	% Advanced maths commands
\usepackage{amssymb}	% Extra maths symbols

\usepackage[T1]{fontenc}
\usepackage{ae,aecompl}

\usepackage{graphicx}	% Including figure files
\usepackage{amsmath}	% Advanced maths commands
\usepackage{amssymb}	% Extra maths symbols

\usepackage{deluxetable}

\renewcommand {\deg}   {\mbox{$^\circ$}}

\newcommand   {\kms}   {\mbox{km\,s$^{-1}$}}
\renewcommand {\ga}    {\mbox{\rlap{\hbox{\lower5pt\hbox{$\sim$}}}\hbox{$>$}}}
\renewcommand {\la}    {\mbox{\rlap{\hbox{\lower5pt\hbox{$\sim$}}}\hbox{$<$}}}

\title[Sgr B2]{Detection of large-scale synchrotron radiation from the molecular\\
        envelope of the  Sgr B  cloud complex at  the Galactic center}

\author[F. Yusef-Zadeh, M. Wardle, R. Arendt, J. W. Hewitt, Y. Hu, A. Lazarian, N. Kassim,\\
 S. Hyman, \& I. Heywood]
{F. Yusef-Zadeh$^1$\thanks{E-mail: zadeh@northwestern.edu}, M. Wardle$^{2}$, R. Arendt$^3$, J. Hewitt$^4$, Y. Hu$^5$, A. Lazarian$^5$, 
\newauthor N. E. Kassim$^6$, S. Hyman$^7$, \& I. Heywood$^{8,9}$\\ 
$^{1}$Department of Physics and Astronomy Northwestern University, Evanston, IL 60208\\
$^{2}$School of Mathematical and Physical Sciences,  Centre for Astrophysics\\
 and Space Technology, Macquarie University, Sydney NSW 2109, Australia\\
$^{3}$UMBC/GSFC/CRESST 2, Code 665, NASA/GSFC, 8800 Greenbelt Rd, Greenbelt MD 20771\\
$^{4}$University of North Florida, Department of Physics, 1 UNF Drive, Jacksonville, FL 32224 , USA\\
$^{5}$Department of Astronomy, University of Wisconsin-Madison, Madison, WI, 53706, USA\\
$^{6}$Code 7213, Remote Sensing Division, U. S. Naval Research Laboratory, 4555 Overlook Ave. SW, Washington, DC 20375, USA\\
$^7$Dept. of Engineering and Physics, Sweet Briar College, Sweet Briar, VA, 24595, USA\\
$^{8}$Astrophysics, Department of Physics, University of Oxford, Keble Road, Oxford, OX1 3RH, UK\\
$^{9}$Department of Physics and Electronics, Rhodes University, PO Box 94, Makhanda, 6140, South Africa\\
}
\date{Accepted XXX. Received YYY; in original form ZZZ}

\pubyear{2019}
\begin{document}
\label{firstpage}
\pagerange{\pageref{firstpage}--\pageref{lastpage}}
\maketitle

%%%%%%%%%%%%%%%%%%%%%%%%%%%%%%%%%%%%%%%%%%%%%%%%%%%%%%%%
\def\msol{\hbox{$\hbox{M}_\odot$}}
\def\lsol{\hbox{$\hbox{L}_\odot$}}
\def\kms{km s$^{-1}$}
\def\Blos{B$_{\rm los}$}
\def\etal   {{\it et al.}}                     % et al
\def\psec           {$.\negthinspace^{s}$}
\def\pasec          {$.\negthinspace^{\prime\prime}$}
\def\pdeg           {$.\kern-.25em ^{^\circ}$}
\def\degree{\ifmmode{^\circ} \else{$^\circ$}\fi}
\def\ut #1 #2 { \, \textrm{#1}^{#2}} % \ut unit p  unit^p 
\def\u #1 { \, \textrm{#1}}          % \u unit     unit
\def\nH {n_\mathrm{H}}
\def\ddeg   {\hbox{$.\!\!^\circ$}}              % Degrees over dot
\def\deg    {$^{\circ}$}                        % Degrees symbol
\def\le     {$\leq$}                            % <=
\def\sec    {$^{\rm s}$}                        % Second of time
\def\msol   {\hbox{$M_\odot$}}                  % Solar mass
\def\i      {\hbox{\it I}}                      % italic I
\def\v      {\hbox{\it V}}                      % italic V
\def\dasec  {\hbox{$.\!\!^{\prime\prime}$}}     % Arcseconds over dot
\def\asec   {$^{\prime\prime}$}                 % Arcseconds symbol
\def\dasec  {\hbox{$.\!\!^{\prime\prime}$}}     % Arcseconds over dot
\def\dsec   {\hbox{$.\!\!^{\rm s}$}}            % Second over dot
\def\min    {$^{\rm m}$}                        % Minutes of time
\def\hour   {$^{\rm h}$}                        % Hours of time
\def\amin   {$^{\prime}$}                       % Arcminutes symbol
\def\lsol{\, \hbox{$\hbox{L}_\odot$}}
\def\sec    {$^{\rm s}$}                        % Second of time     
\def\etal   {{\it et al.}}                     % et al.
\def\la{\lower.4ex\hbox{$\;\buildrel <\over{\scriptstyle\sim}\;$}}
\def\ga{\lower.4ex\hbox{$\;\buildrel >\over{\scriptstyle\sim}\;$}}
%%%%%%%%%%%%%%%%%%%%%%%%%%%%%%%%%%%%%%%%%%%%%%%%%%%%%%%%

\begin{abstract} 
We present highly sensitive measurements taken with MeerKAT at 1280 MHz as well as archival GBT, MWA and VLA images at 333, 88 and 74 MHz.  We report 
the detection of synchrotron radio emission from the infrared dark cloud (IRDC) associated with the halo of the Sgr B complex on a scale of $\sim$60 
pc. A strong spatial correlation between low-frequency radio continuum emission and dense molecular gas, combined with spectral index measurements, 
indicates enhanced synchrotron emission by cosmic-ray electrons. Correlation of the FeI 6.4 keV K$\alpha$ line and synchrotron emission provides 
compelling evidence that the low energy cosmic-ray electrons are responsible for producing the K$\alpha$ line emission. The observed synchrotron 
emission within the halo of the Sgr B cloud complex has mean spectral index $\alpha\sim-1\pm1$ gives the magnetic field strength $\sim100\mu$G for 
cloud densities $n_\mathrm{H} = 10^4$--$10^5$\,cm$^{-3}$ and estimate  cosmic-ray ionization rates between 10$^{-13}$ and 10$^{-14}$ s$^{-1}$. 
Furthermore, the energy spectrum of primary cosmic-ray electrons is constrained to be $E^{-3\pm1}$ for typical energies of few hundred MeV.  
The extrapolation of this spectrum to higher energies is consistent with X-ray and $\gamma$-ray emission detected from this cloud. These measurements 
have important implications on the role that high cosmic-ray electron fluxes at the Galactic center play in production of radio synchrotron emission, the FeI 
K$\alpha$ line emission at 6.4 keV and $\sim$GeV $\gamma$-ray emission throughout the central molecular zone (CMZ).  
\end{abstract}

%%%%%%%%%%%%%%%%%%%%%%%%%%%%%%%%%%%%%%%%%%%%%%%%%%

%%%%%%%%%%%%%%%%% BODY OF PAPER %%%%%%%%%%%%%%%%%
\section{Introduction}

The nucleus of our Galaxy, centered on the supermassive black hole Sgr A*, is characterized by molecular clouds with high temperatures, high densities, and large turbulent 
linewidths throughout the inner few hundred parsecs, the so-called the Central Molecular Zone (CMZ) \citep{2022arXiv220311223H}. 
H$_3^+$ absorption line observations imply 
high cosmic-ray ionization rates, $\zeta \sim 10^{-14}$ to 10$^{-13}$ s$^{-1}$ throughout the CMZ
at levels a hundred to a thousand times that of the solar neighborhood 
\citep{2005ApJ...632..882O,2012ApJ...745...91I,2016A&A...585A.105L,2019Natur.573..235H,2019MNRAS.490L...1Y,2022ApJ...925..165H},  and
heat  the gas clouds to high temperatures \citep{2007ApJ...665L.123Y}. These observations indicate 
that the cosmic-ray 
pressure in the CMZ is significant when compared to the interstellar gas pressure of the Galactic center.  
The Galactic center is also known to harbor a large number of molecular clouds 
showing  evidence of high energy processes  traced by the K$\alpha$  FeI line at 6.4 keV, as well as GeV and TeV emission 
\citep{2010ApJ...714..732P,2010PhRvD..82d3002A,2013ApJ...762...33Y,2015ApJ...809...48D}. 

Given that the Galactic center is host to a high cosmic-ray flux and dense molecular clouds, this region is an excellent target to search for 
synchrotron radio continuum
emission from the molecular clouds that make up the CMZ
\citep{2005ApJ...632..882O,2022arXiv220311223H}. 
One significant issue involves the relative contributions of low-energy cosmic-ray electrons and protons in producing the ionization and the 6.4 keV 
K$\alpha$ line emission. 
In principle, synchrotron emission from molecular clouds  is a clear  discriminator   
between the two scenarios; high masses of protons emit negligible   synchrotron radiation. 
In addition, if cosmic-ray electrons are responsible for the emission, 
synchrotron radiation places  constraints on the strength of the magnetic field inside dense molecular clouds.
The path length over which cosmic-ray electrons travel is limited 
by the magnetic field geometry, and collisional and
bremsstrahlung losses in the denser medium. Cosmic-ray electrons that do penetrate deep into a cloud will interact with 
the magnetic field of molecular clouds to produce synchrotron radiation. 

%The magnetic field in this situation is stronger in dense clouds than the 
%diffuse interstellar medium, 
%so the synchrotron emission from denser molecular clouds is expected to be stronger than from diffuse clouds.  

%The production of ionized and K$\alpha$ emission by low-energy cosmic-ray particles are 
%There are two scenarios that Low energy cosmic-ray electrons and protons

The magnetic field strength in interstellar molecular clouds provides a critical role in the process of star formation. Furthermore, the energy spectrum of
cosmic-ray particles within   molecular clouds plays a crucial role in the physics and chemistry of dense gas. The magnetic field within 
molecular clouds has  
been  probed using challenging
techniques such as Zeeman splitting and, optical, near-infrared and sub-millimeter polarization 
studies \citep{1996ApJ...462L..79C,2011AJ....142...33A,2011A&A...529A..95V,2018A&A...616A..56A}. 
Synchrotron emission  depends
only on the magnetic field strength and the spectrum of  cosmic-ray electrons at energies of a few hundred MeV. 
 This method relies on 
few hundred MeV  cosmic-ray 
particles that diffuse from outside into molecular 
clouds. 
Theoretical prediction of synchrotron  emission from molecular clouds was made decades
ago \citep{1977ApJ...212..659B,2018A&A...620L...4P,2022A&ARv..30....4G,2018A&A...616A..56A}. 
Another prediction to determine the magnetic field orientation 
 uses the aniosotropic property of MHD turbulence 
to trace the orientation of the magnetic field within clouds \citep{2017ApJ...842...30L}. 

%, but as yet, there has not been a convincing detection of synchrotron emission from
%molecular clouds \citep{2014arXiv1412.4500S,2013MNRAS.436.2127O,2011AJ....141...82J}.  

A number of attempts have been made in the past to detect nonthermal radio continuum emission
from molecular clouds \citep{2014arXiv1412.4500S,2013MNRAS.436.2127O,2011AJ....141...82J,2008MNRAS.390..683P,2018A&A...616A..56A}. However, these 
searches suffered from
limited sensitivity, contamination by strong free-free thermal sources
 and a lack of low-frequency observations 
\citep{2005ApJ...632..882O,2022arXiv220311223H}.  
Low-frequency observations are helpful  
because optically thin synchrotron emission rises at ,at frequencies  
 whereas thermal sources follow a blackbody spectrum, 
 with intensity decreasing at low frequencies,
$I_\nu \propto \nu^2$,  and are   observed in absorption 
suppressing background emission and appearing as holes in 74 MHz images 
\citep{2006AJ....132..242N}.

%The Sgr B complex  shows  a wide range of structures such as ultracompact HII regions, molecular cores  and envelopes surrounded by a halo of molecular gas
%\citep{1990ApJ...350..186G}. 

As a representative molecular cloud, the  Galactic center cloud, G0.13-0.13,  was studied 
in the CMZ  showing evidence   steep-spectrum synchrotron emission  at 74 MHz \citep{2007ApJ...656..847Y,2013JPCA..117.9404Y}.  
We focus here on the Sgr B complex, a prominent star forming molecular cloud in the CMZ, unlike the 
G0.13-0.13 cloud which shows no trace of star formation activity \citep{2013JPCA..117.9404Y}.  
The inner region of the Sgr B cloud is  one of the 
most luminous,  massive star-forming regions in the Galaxy 
containing hot molecular cores, a large concentration of ultracompact HII regions, and diffuse free-free emission from ionized gas \citep{1991ApJ...369..157L,2014ApJ...781L..36D}.
A number of studies indicated that synchrotron  emission arises from Sgr B2 
\citep{2005ApJ...626L..23L,2007ApJ...665L.123Y,2007ApJ...660L.125H,2008MNRAS.390..683P,2011AJ....141...82J,2016ApJ...819L..35Y,2019A&A...630A..73M,2022FrASS...9.9288R}.
In addition, an OH(1720 MHz) maser was detected 
from a 150 MHz continuum source Sgr B2(M), suggesting  cosmic-rays  heat  Galactic center molecular clouds  in 
the  Central molecular Zone (CMZ) \citep{2016ApJ...819L..35Y}. 
Most previous studies  have focused on the interior of the Sgr B cloud complex, 
but here we study the  outer envelope,  
a hook-shaped halo structure roughly $25'\times10'$ in extent  surrounding Sgr B1 and Sgr B2,  from which we report nonthermal radio continuum 
emission. 

%We apply a method that has hardly been explored in the past to estimate the magnetic field strength in the interior of the giant molecular clouds Sgr B in the Galactic 
%center.

%there are no obvious thermal radio continuum compact sources. 
 
\section{Past Measurements} 

Details of the VLA and MeerKAT observations at 74, 333, and 1280 MHz  presented here have been described previously 
\citep{2003ANS...324...17B,2004AJ....128.1646N,2022ApJ...925..165H}. These observations have two advantages when compared to previous measurements.  
First, MeerKAT observations provided a sensitive mosaic of the inner few degrees of the Galactic center at 1280 MHz 
(with rms noise $\sigma\sim 80 \mu$Jy per 4$''\times4''$ beam) revealing numerous structures that had not been seen in previous radio surveys. This 
survey was able
to measure  the spectral index ($\alpha$), where the intensity $I_\nu \propto \nu^\alpha$, using 16 sub-bands between 
856 and 1712 MHz. Second, low frequency observations of the Galactic center were carried out with the VLA at 74 and 333 MHz (4m and 0.9m wavelengths) 
and with the GBT at 333 MHz 
\citep{2003ANS...324...17B,2005ApJ...626L..23L,2004AJ....128.1646N}
 to separate nonthermal and thermal sources. 
The original 330 MHz image was made by “feathering” 
together multi-configuration VLA with GBT 
data, as described
 in \citep{2017PASP..129i4501C}.  The image was subsequently smoothed in AIPS using the tasks 
CONVL and OHGEO to make a lower resolution image.  

Murchison Widefield Array (MWA) observations of the Galactic plane  were made at multiple frequencies.  We selected the lowest frequency image at 88 MHz 
toward Sgr B having a resolution of $\sim328''\times310''$ \citep{2022PASA...39...35H}. The image  shows a very similar  
morphology to  that of the VLA image at 74 MHz.

\section{Results} 

\subsection{Morphology and spectral index} 

The extreme column densities in Sgr B2 make it optically thick at mid-IR wavelengths, thus it appears 
as an IRDC superimposed on a brighter background, as shown in the  8$\mu$m Spitzer image in Figure 1a. 
The envelope of the cloud 
resembles a hook  with a length of $\sim25'$ elongated along
the Galactic plane, and a circular structure to the SE with diameter  of $\sim12'$.  The broken outline  is motivated by diffuse 
emission 
in the 1280 MHz (20cm)
image (below). The thickness of the elongated feature (hearafter $IRDC\, hook$  or $hook$) 
ranges between $\sim3'-6'$,  or $\sim7-17$ parsecs at 
the Galactic center distance of 8.0 kpc. 
The outline avoids the bright
regions of star formation activity traced by the evolved HII region Sgr B1, and compact and ultracompact HII regions Sgr B2.

The column density of the IRDC as traced by far-IR emission from cold dust is shown in Figure 1b, which clearly reveals the parent dense cloud within which 
the HII regions are
embedded \citep{2011ApJ...735L..33M}. Typical hydrogen column densities, $N_{\rm H}$,  are in the range of a few times 
$10^{23.5}-10^{25.5}$ cm$^{-2}$. 
Figure 1c shows radio continuum
emission at 1280 MHz where we note faint diffuse emission distributed throughout the Sgr B IRDC \citep{2022ApJ...925..165H}. Typical intensities within the IRDC hook are
$\sim1-2\times10^{-4}$ Jy beam$^{-1}$, about 4-5  times stronger than the background emission. Although the continuum emission from the bright Sgr B complex outside the IRDC hook
is dominated by thermal continuum emission, the outline of 1280 MHz emission appears to correlate spatially with the IRDC. Figures 1d,e,f  show 
333 MHz, 88 MHz and 74 MHz counterparts to 1280 MHz
emission. Diffuse emission is still seen throughout 
the IRDC hook, whereas the thermal emission of Sgr B1 and B2 becomes optically thick, and these regions are now seen in
absorption at $<100$ MHz.  This implies that the emission coincident with the IRDC hook is synchrotron radiation.

Figure 1a-d reveal a 
spatial correlation between nonthermal radio continuum emission and dense molecular clouds, supporting the idea that enhanced synchrotron emission
results from cosmic-ray electrons. 
Further support for synchrotron nature of this emission  comes from the 
spectral index. Figure 2a shows the 
spectral
index measured across the 1280 MHz band.  To the SE, the hook has a mean spectral index of $\alpha\sim -0.75\pm0.25$ 
consistent with  synchrotron radiation resulting from cosmic-ray interaction. 
 Figure 2b 
shows the spectral
index distribution between 333 and 74 MHz. The mean spectral index is $-0.75$ toward the SE and $-0.5$ toward the linear segment of the hook, north of thermal
continuum emission  associated with Sgr B1 G0.5-0.0 and Sgr B2 G0.65-0.0. The flatter  spectrum  suggests that the emission may be  contaminated by 
free-free
emission  from Sgr B. 
Low-frequency intensities at 74 and 333 MHz suffer from ionospheric-related  calibration, uncertainties in 
 different spatial frequency coverage, 
 we give a spectral 
index  $\pm0.5$ in the Sgr B halo  due to systematic errors. In addition,
the contamination of free-free emission  at 333 and 1280 MHz, we estimate flux values as  upper limits. 
After accounting for systematic errors due to contamination of free-free emission, 
the spectral index is  estimated to be  $\alpha\sim-1\pm1$.

\subsection{Magnetic field geometry}
 
%The intensity maps above indicate non-thermal values of the spectral index for the hook, but the result can also be sensitive to the background level. 
%To confirm the nonthermal values, we selected locations with 
%customized background subtraction. We estimated the spectral index using 1D slices across the intensity maps.  Figure 2c shows that the spectral index values are [-1.4, -0.3] for slice 1 and 
%[-1.7, -1.5] for slice 2. The spectral index estimates combined with a correlation of the spatial distribution of molecular clouds and radio continuum emission from the Sgr B complex provide the 
%evidence for synchrotron emission from a large-scale molecular cloud.  We find an average spectral index of $-0.75\pm0.25$ which is more consistent with 20cm data.

There are no radio continuum polarization measurements of the Sgr B cloud, but we can use a relatively new technique to determine the geometry of the 
magnetic 
field within the hook, the Synchrotron Intensity Gradients (SIGs) technique (\citealt{2017ApJ...842...30L,2023arXiv230610011H}). 
SIGs is based on the anisotropic nature of magnetohydrodynamic (MHD) turbulence, where the maximum velocity and magnetic field fluctuations
occur perpendicular to the magnetic field direction, and turbulent eddies are  elongating along the magnetic field.
This anisotropy has been demonstrated through MHD simulations 
\citep{2000ApJ...539..273C,2002ApJ...564..291C,2022MNRAS.512.2111H} and  as in-situ measurements in the solar wind 
\citep{2020FrASS...7...83M,2021ApJ...915L...8D}. Since synchrotron radiation arises from relativistic electrons moving perpendicular to the magnetic 
field lines \citep{1947PhRv...71..829E,1979rpa..book.....R}, this anisotropic property extends to synchrotron emission as well 
\citep{2012ApJ...747....5L}. 
Consequently, the synchrotron intensity gradients rotated by $\pi/2$  degrees and can be used to infer the orientation of the 
magnetic field.

The input of SIGs is a intensity map of synchrotron emission \textbf{$I(x,y)$}. We adopt the same procedure as  
\cite{2023arXiv230610011H}. The intensity map is convolved with 3 $\times$ 3 Sobel kernels to calculate a pixelized gradient map $\psi(x,y)$. The measurement of an 
individual 
gradient, denoted as $\psi(x,y)$, however, does not provide an accurate representation of the magnetic field direction. Anisotropy within turbulence is a statistical 
concept that becomes apparent only when a sufficient number of samples are considered. To identify this anisotropy, we employ the following approach: (1) The gradient angle 
within each sub-block (20x20 pixel$^2$) is statistically computed using the sub-block average method. (2) The resulting sub-block averaged gradient map, denoted as 
$\psi_{\rm s}(x,y)$, is then utilized to construct pseudo Stokes parameters, specifically $Q_{\rm g} = I(x,y)\cos(2\psi_{\rm s}(x,y))$ and $U_{\rm g} = 
I(x,y)\sin(2\psi_{\rm s}(x,y))$. To reduce noise, we apply a masking procedure to exclude gradients with corresponding intensities lower than $1\times10^{-4}$ Jy/beam. 
Additionally, the $Q_g$ and $U_g$ parameters are smoothed using a Gaussian filter, with the full width at half maximum (FWHM) equal to the size of the sub-block. The 
SIGs-derived magnetic field angle is then given by: $\psi_{\rm B}=\psi_s(x,y)+ \pi/2$.
Figure 2c shows the magnetic field geometry of this region suggesting that the magnetic 
field is not parallel to the edge of the hook, allowing cosmic rays to penetrate deeper into the cloud. 

%In Alfv\'en MHD turbulence, the symmetry of velocity and magnetic 
%fluctuations leads to the maximum magnetic field fluctuation also aligning with the perpendicular direction. 

\section{Discussion}

Here we use the observed synchrotron intensity  to infer the cosmic-ray electron population present with the IRDC hook
and estimate the magnetic field strength in the Sgr B halo,  and the ionization rate and Fe K$\alpha$ 
line emission due to these interactions within the Sgr B halo.   We do this by assuming that the cosmic-ray electron flux incident on the cloud has a 
power-law energy spectrum, and compute  the 
modified spectrum within the halo accounting for energy losses due to interactions with the cloud material (see below). 
  The incident cosmic-ray 
electrons traverse a variety of paths 
depending on the magnetic field geometry, but for simplicity, we assume that the electrons have traced 
a single equivalent hydrogen column, $N_\mathrm{H}$.  
For an adopted magnetic field strength in the halo,  $B$, the normalization and power-law index of the incident spectrum is adjusted so that the 
loss-modified electron spectrum is 
consistent with the 74\,MHz synchrotron intensity and 74\,--\,333\,MHz spectral index.  We then compute the implied cosmic-ray ionization rate and 
the FeI 
K$\alpha$ 
line intensity,  finding overall consistency if $N_\mathrm{H}\sim10^{24}$\,cm$^{-2}$ and $B\approx100\,\mu$G. Finally, we place additional 
constraints  on the energy spectrum of cosmic-rays when extrapolated to high energies and compare  predicted and observed $\gamma$-ray  measurements.  

% Check the 1300 MHz intensity - this will be primarily free-free?

\subsection{Cosmic-ray electron energy loss vs column density}

At 74\,MHz the flux density in the bright region to the southeast of the hook is 4.63\,Jy 
integrated over a 4.14$'\times8'$ region. After background subtraction, the flux density 
is 2.25 Jy.  At 333\,MHz the measured background-subtracted intensity over the same region 
is an upper limit on the synchrotron intensity because of contamination by free-free 
emission.  We detect a background-subtracted synchrotron intensity of $\sim 0.5$\,Jy after 
one Jy beam$^{-1}$, after using  a nearby $off$ region to estimate the background, thus 
 yielding a synchrotron spectral index $\alpha=-1$.
 Adopting 10\,pc as the source depth along the line-of-sight, we infer  the 74-MHz synchrotron volume emissivity 
to  be  $j_\nu\approx 
3.1\times10^{-36}$\,erg\,cm$^{-3}$\,s$^{-1}$\,Hz$^{-1}$\,sr$^{-1}$, with $j_\nu\propto \nu^{-1}$.

What does this tell us about the cosmic-ray electron population? For a given magnetic field strength the emissivity is primarily determined by the 
differential number density per unit energy of electrons $n(E)$ at energies of order
\begin{equation}
    E_\nu \;=\;  \left(\frac{4\pi m_e c \nu}{3eB}\right)^{1/2} \!\!m_e c^2 \;\approx\; 215 \left(\frac{\nu_{74}}{B_{100}}\right)^{1/2}\;\mathrm{MeV}\,.
    \label{eqn:Enu}
\end{equation}
where $\nu_{74} \equiv \nu/ (74 {\rm MHz})$ and  $B_{100} \equiv B/ (100 \mu\rm{G})$. 
The synchrotron emissivity is 
\begin{equation}
        j_\nu \;=\; \frac{2}{3}\;\frac{\sqrt{3}}{4\pi}  \, \frac{e^3B}{m_ec^2}\; f\, E_\nu\,n(E_\nu)  
    \label{eqn:jnu}
\end{equation}
where the spectral form factor $f$ is of order unity as long as the electron spectrum does not have significant structure 
within a factor of a few of $E_\nu$.
For simplicity 
we have assumed a tangled field and that the electron
 pitch-angle distribution is isotropic.  
 The logarithmic slope of 
the electron spectrum in the neighborhood of $E_\nu$ is approximately related to 
the spectral index $\alpha$ of the synchrotron emission, $\alpha$ defined in the 
usual way, i.e. $d\log n/d\log E \approx 2\alpha-1 = -3$. 
Setting $f=1$, $B=100\,\mu$G and adopting the 74\,MHz emissivity estimated above, we find $E\,n(E) \approx 2.5\times10^{-9}$\,cm$^{-3}$ at 215\,MeV.

While the synchrotron emission enables good characterization of the electron spectrum at a few hundred MeV,  it is the interactions of the more 
numerous electrons in the 
MeV energy range that are responsible for the bulk of heating, ionization, and  Fe K$\alpha$ emission.   To 
 model the electron spectrum at lower energies, we assume that 
the cosmic-ray electrons incident on the molecular cloud have a power-law spectrum,  and then consider how this is modified by collisional 
and bremsstrahlung losses within 
the cloud.   For kinetic energies in the range 0.01\,MeV -- 1\,GeV we use the tabulations
of electron stopping power and range  provided by the NIST ESTAR 
database \citep{1982spre.reptR....B}  for a 5:1 by number mixture of molecular hydrogen and helium.  At energies above 1\,GeV we 
use the approximate analytic 
expressions for ionization and brehmstrahlung losses provided by \cite{1976PhRvA..13.1278Q}
 adjusted by a few percent 
to match the NIST tabulations at 1\,GeV. 
Figure 3a shows the hydrogen column that can be traversed by electrons before losing their energy\footnote{or technically running out of $steam$}.  We 
see that incident electrons with energies below 
1\,MeV lose all their 
energy before traversing $10^{23}$\,cm$^{-2}$, while those with $E>10\,$MeV have ranges exceeding $10^{24}$\,cm$^{-2}$ or more \citep{2015ApJ...809...48D}. 
Note that this implies that  our 
results are  insensitive to the adopted incident electron spectrum below a few MeV.

%This provides the collisional and radiative stopping powers of the medium $S(E) = dE/d\lambda$ .

\subsection{Cosmic-ray electron spectrum vs column density}

Following \cite{2012ApJ...756..157G}, we compute  the electron spectrum in the cloud, assuming for simplicity that the electrons have all traversed 
the same 
column density $N_\mathrm{H}$.  Then for a given field strength, the power-law index $p$ and normalization of 
the incident electron spectrum, $n(E)\propto\, E^{-p}$,  are adjusted so that the 
synchrotron emissivity of the attenuated electron population yields the inferred 74 MHz emissivity and 74-333\,MHz spectral index (i.e. $\alpha = -1$).  The synchrotron 
emissivity is computed by explicitly integrating over the electron spectrum, using the  approximation for the synchrotron emissivity 
given in equation D7  
of \cite{2010PhRvD..82d3002A}.

Figure 3b shows the resulting electron energy spectra for $B=100\,\mu$G and $N_H = 10^{23}$--$10^{25}$cm$^{-2}$.  
Here we have plotted $En(E)$ rather than $n(E)$, as this is proportional to electron number density per logarithmic energy interval.  
The normalization and slope of the 
electron spectra at $E_\nu\sim215$\,MeV, are  very close to the  estimate made above.  
This  is  a direct consequence of the 
requirement to 
match the 
synchrotron emissivity at 74\,MHz and the  spectral index between 74\,MHz and 333\,MHz.  The spectra are barely attenuated above 1\,GeV. 
Energy losses 
are minimal at  lower energies, as $n(E)$ flattens and rolls over as the energy lost in traversing the given column becomes 
significant 
(c.f.~Figure 3c). For  $N_\mathrm{H}\sim10^{25}$\,cm$^{-2}$, the roll over begins at 1\,GeV, so the incident spectrum must be 
steeper  than at higher energies to 
compensate for flattening of the spectrum at $E_\nu$ due to the induced ionization loss.

\subsection{Cosmic-ray ionization rate vs the magnetic field}

We integrate over the electron spectrum to determine the energy loss rate per unit volume due to 
collisional interactions, and convert this to ionization by noting that on 
average an ionization occurs for each $w=$40.1\,eV lost in collisional interactions 
\citep{1999ApJS..125..237D}, i.e., $\zeta= (1/w\, n_H)\, \int_0^\infty{n(E)\, L(E)\, dE} $  
where $L(E)$ is the energy loss rate due to collisional interactions
of a cosmic-ray electron of energy $E$ with ambient molecular gas.  
The inferred ionization rate is plotted against $B$ in
Figure~3c.  The steep decline in ionization rate with increasing field strength occurs because $E_\nu \propto B^{-1/2}$ while $E_\nu n(E_\nu)$ must 
scale   as $B^{-1}$ to match the 74-MHz emissivity (see eqs~\ref{eqn:Enu}--\ref{eqn:jnu}).  The net effect is to slide the electron spectra in Figure 
3b  to the left and downwards, lowering the rate of 
energy deposition and hence the ionization rate.

%XXThe equipartition field strength neglecting any contribution from protons is 81\,$\mu$G; this is likely a lower bound on $B$.

We see that for the typical column density in the halo of Sgr B, $N_\mathrm{H}\sim10^{24}$\,cm$^{-2}$, the 
inferred 
ionization rate is consistent with the extreme rates , i.e.~$10^{-14}$--$10^{-13}$\,s$^{-1}$\,H$^{-1}$,  
inferred in the CMZ from H$_3^+$ absorption line observations. 
In this case,  the magnetic field is of order 
$100\,\mu$G, roughly in line 
with the typical field measured in molecular gas at the $n_\mathrm{H} = 10^4$--$10^5$\,cm$^{-3}$ densities implied by the 10\,pc 
sightline through the Sgr B2 halo.

\subsection{FeI K$\alpha$ line at 6.4\, keV vs magnetic field}

As the observed synchrotron emission at 74 MHz arises from electrons with energies of a few 100 MeV, the extrapolation to 10 MeV is rather uncertain. 
This can be probed by considering the impact of low energy cosmic-ray particles with neutral gas which is expected to produce 6.4 keV FeI K$\alpha$ 
line emission. In fact, a number of past measurements have reported localized 6.4 keV line and emission from the cosmic-ray  enhanced 
molecular ion such as PO$^+$  from the Sgr B 
complex \citep{2019A&A...630A..73M,2022ApJ...934...19R,2022FrASS...9.9288R}. 
The 6.4 keV line emission has been interpreted as arising either  
 by  bombardment of low energy cosmic-ray particles onto a molecular cloud, or X-ray reflection from a powerful flare from Sgr A* 
that occurred a few hundred years 
ago \citep{2010ApJ...714..732P,2013ApJ...762...33Y}.  The equivalent width (EW) and intensity of the 6.4 keV K${\alpha}$ line emission from the Sgr B 
complex,  derived from Chandra and Suzaku observations,  are presented in Figure~4a,b respectively.  The EW of the diffuse region to the SE is 
270$\pm47$ eV.  
The X-ray emission from the SE of the hook correlates well with nonthermal radio continuum and molecular gas, as shown in Figure 1. The low value of 
the EW, indicating K-shell ionization  by low-energy cosmic-ray electrons because cosmic-ray electrons also produce a nonthermal brehmstrahlung 
continuum.  This  
combined with the spatial correlation between low-frequency radio and X-ray 
images strongly suggests that the picture that the 6.4 keV line emission results from the interaction of sub-relativistic cosmic-ray electrons with 
energies 
of few 10-100 
keV and the molecular gas in the interior of the Sgr B complex.

Cosmic-ray electrons  produce 6.4 keV iron K$\alpha$ photons via the collisional ejection of K-shell electrons from neutral iron atoms, followed 
by emission 
of a photon as the shell vacancy is filled by an electron in a higher shell \citep{2000ApJ...543..733V}.  We compute the line-integrated intensity of 
the K$\alpha$ emission 
using the parameterization of the K-shell ionization 
cross-section in \citep{1976PhRvA..13.1278Q}, and correcting for the fraction of K-shell ionizations that produce a K$\alpha$ photon, adopting an iron 
abundance of twice solar, i.e. 
$n(\mathrm{Fe})/n_\mathrm{H} = 5.6\times10^{-5}$, consistent with the enhanced 
metalicity at the Galactic center.  The  K$\alpha$ intensity 
as a function of 
magnetic field strength is plotted in Figure 3c. As with the ionization rate, 
there is a steep decline in  line flux with increasing magnetic field.  
he  K$\alpha$ intensity is not as sensitive to $N_\mathrm{H}$ as the ionization rate,  because  
the  attenuation 
of the cosmic-ray electrons with increasing column density is partially offset by the increase in the number of target iron atoms along the line 
of sight. 
  The typical K$\alpha$ flux from the Sgr B halo, 
$3\times10^{-6}$\,ph\,s$^{-1}$\,arcmin$^{-2}$ (see Fig. 3a)  is indicated by the horizontal dotted line, and is consistent with 
the hydrogen column density of $\sim10^{24}$ cm$^{-2}$, if B$\sim 100\,\mu$G.

\subsection{Energy spectrum constraints by  $\gamma$-ray emission}

Additional constraints can be obtained by extrapolating  the energy spectrum of electrons  to high energies.
The radiation from relativistic electrons was computed using {\it naima} \citep{naima}, and adjusted to fit the radio flux and spectral index observed
within  the MeerKAT broad band at 20cm and between 74 and 333 MHz. This results  in medium-energy $\gamma$-ray bremsstrahlung emission that 
strongly 
peaks at photon 
energies of 100 MeV, 
as shown in Figure 4d 
\citep{2021ApJS..256...12A}.
An upper-limit on the $\gamma$-ray flux at 100 MeV places a lower-limit on the magnetic field
strength, assuming the target density (${\rm n = 10^{3}-10^{4} \, cm^{-3}}$) assuming a depth 10pc and particle
spectrum ($E^{-3}-E^{-4}$).

Fermi-LAT is unable to resolve an extended source, 
 particularly given the high Galactic foreground towards the Galactic center.
 However, Fermi-LAT 
is capable of detecting an excess  differential flux $E^2 \frac{dN}{dE}$ of $10^{-11}$ erg cm$^{-2}$ s$^{-1}$ in a $3^\circ$ field.
 No source is detected spatially 
coincident with recently created 
synchrotron nebula in the 12-year Fermi-LAT source catalog \citep{4fgl-dr3}.
The 68\% containment angle even for just the top quartile of LAT events at 100 MeV is 3 degrees,
making it difficult to separate this emission component from other sources at the Galactic center.
The predicted bremsstrahlung $\gamma$-rays have a unique spectral signature, which may allow the Sgr B source   to be distinguished from other Galactic center $\gamma$-ray sources
\citep{2013ApJ...762...33Y}. It is clear that 100 $\mu$G magnetic field  exceeds the detection threshold.
For a target density of $10^4$ cm$^{-3}$ and a magnetic field of 80 $\mu$G the predicted $\gamma$-ray flux at 100 MeV is two times brighter than the $\gamma$-ray source associated with
Sgr A (4FGL J1745.6-2859). Thus, lower (higher) magnetic field strengths of higher (lower) target densities are ruled out by Fermi-LAT observations.

\section{Summary}

The halo of the molecular emission from the prominent giant molecular complex, Sgr B2 in the CMZ, coincides with radio continuum
synchrotron emission. The spatial correlation between emission at radio, IR, and
X-ray wavelengths, provided compelling evidence that 
relativistic electrons are diffusing into the cloud from the outside.  A  quiescent molecular cloud, G0.13-0.13 
in the CMZ,  also shows  evidence of nonthermal continuum emission. These two molecular clouds,  Sgr B and G0.13-0.13, 
support the spatial correlation of the 74 MHz emission and molecular clouds in the Galactic center
\citep{2013JPCA..117.9404Y}. 
  Thus,  clouds in the CMZ   should all emit  nonthermal radio emission because of 
high cosmic-ray ionization rate noted throughout the CMZ. 
The high cosmic-ray electron 
ionization 
rate in 
dense clouds with densities $\sim10^{4-5}$ cm$^{-3}$ give the magnetic field strength of $100\mu$G as the electrons diffuse into the CMZ clouds and 
heat 
the gas clouds to high temperatures. 
Lastly, the evidence for the FeI K$\alpha$ line emission at 6.4 keV and nonthermal radio continuum emission provide 
a compelling evidence that low-energy cosmic-ray electrons and not protons are responsible for the observed X-ray emission from the  Sgr B cloud  in 
the Galactic center.

\section{Data Availability}

All the data including  VLA and MeerKAT  that we used here are available online and are not proprietary.
We have reduced and calibrated these data,  and these products are available if  requested.

\section*{Acknowledgments}

We thank D. Kaplan for providing initial MWA images of the Galactic center. 
This work is partially supported by the grant AST-2305857 from the NSF.
 Work by R.G.A. was supported by NASA under award No. 80GSFC21M0002. Basic research in radio astronomy 
at the Naval Research Laboratory is funded by 6.1 Base funding.

%This paper makes use of the following ALMA data: ADS/JAO.ALMA\#2011.0.00005.SV. ALMA is a
%partnership of ESO (representing its member states), NSF (USA) and NINS (Japan), together with NRC (Canada) and NSC and
%ASIAA (Taiwan), in cooperation with the Republic of Chile. The Joint ALMA Observatory is operated by ESO, AUI/NRAO and
%NAOJ.

%%%%%%%%%%%%%%%%%%%%%%%%%%%%%%%%%%%%%%%%%%%%%%%%%%


%\vfil\eject

%\vfill\eject
\clearpage

\bibliographystyle{mnras}
\bibliography{example.bib}

\clearpage

\begin{figure}
%\epsscale{0.2}
\centering
%\hspace{1.0in}
%\vspace{-0.2in}
   \includegraphics[scale=0.3,angle=90]{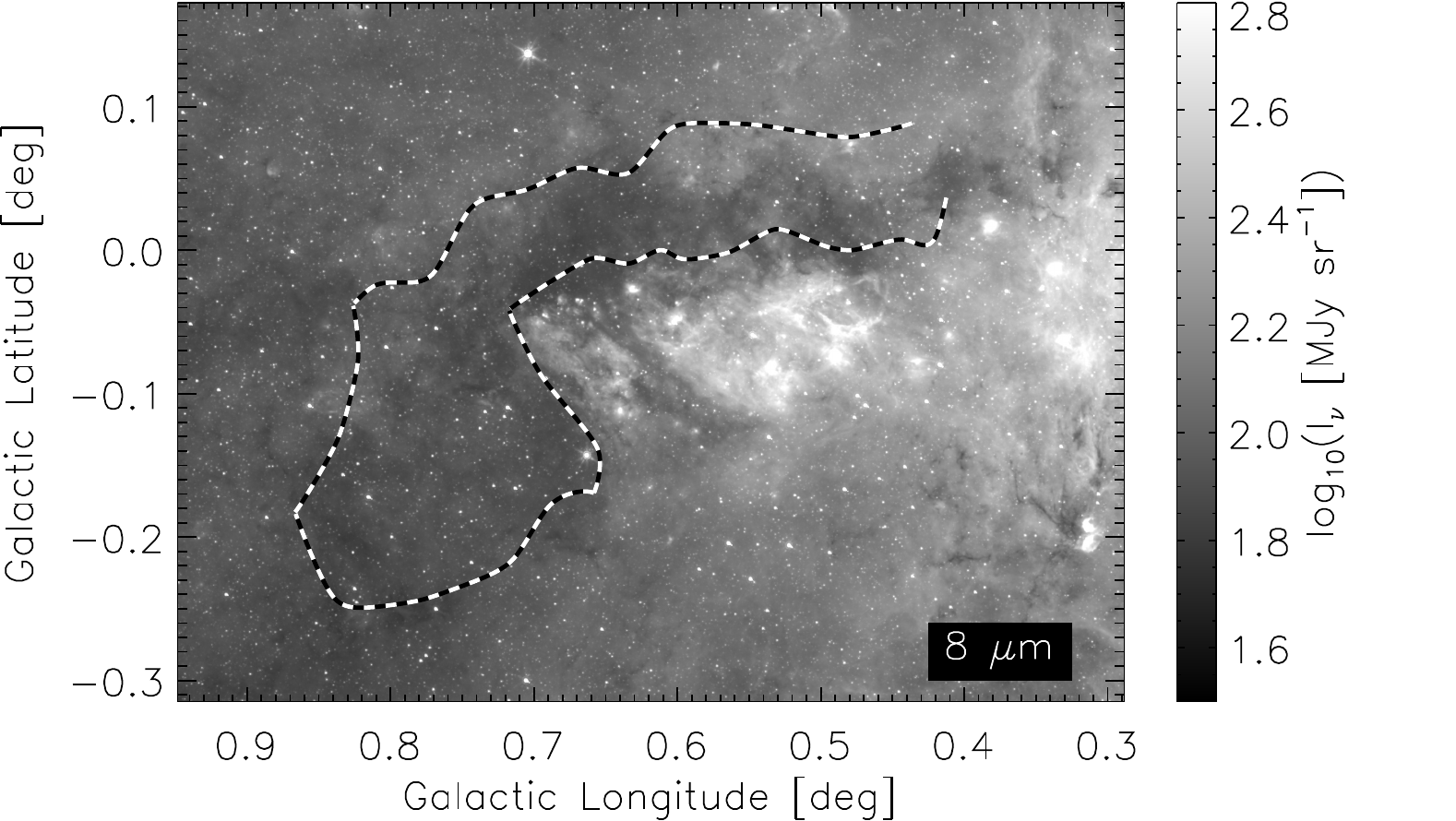}
 \hspace{0.5in}  
    \vspace{0.25in} 
  \includegraphics[scale=0.3,angle=90]{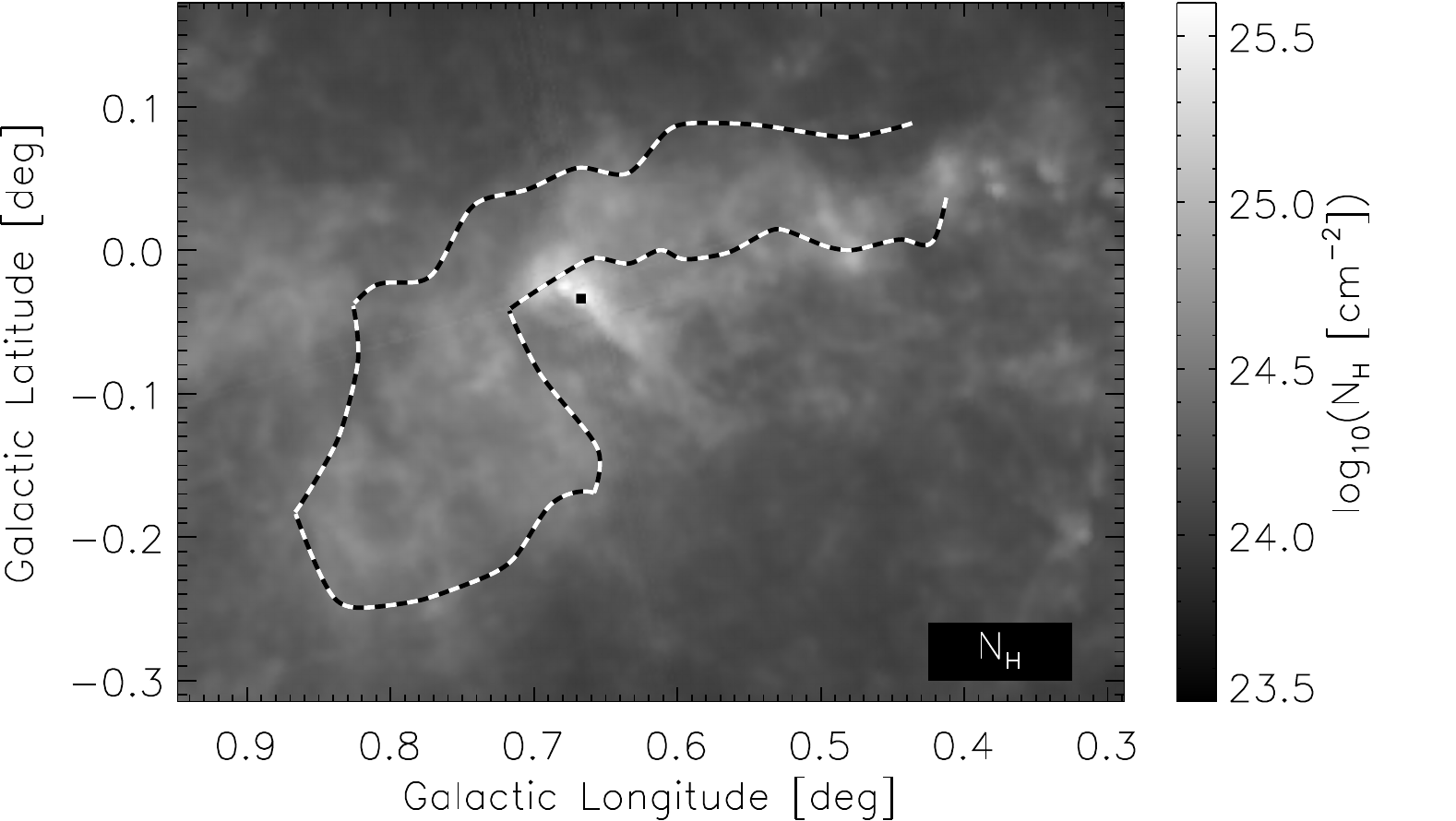}
 \vspace{0.25in}    
\includegraphics[scale=0.3,angle=90]{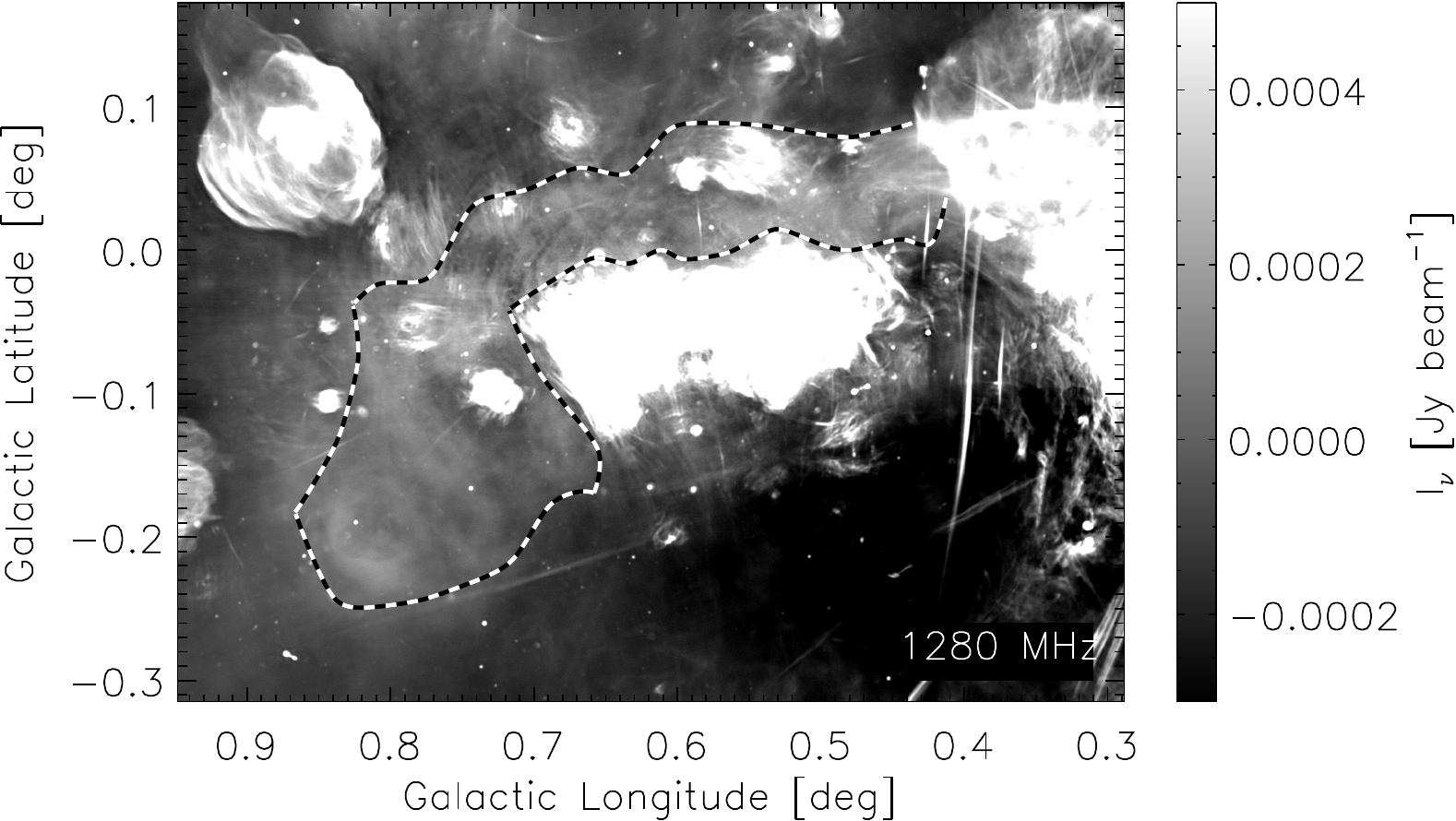}
%    \vspace{0.25in} 
    \hspace{0.5in} 
\includegraphics[scale=0.3,angle=90]{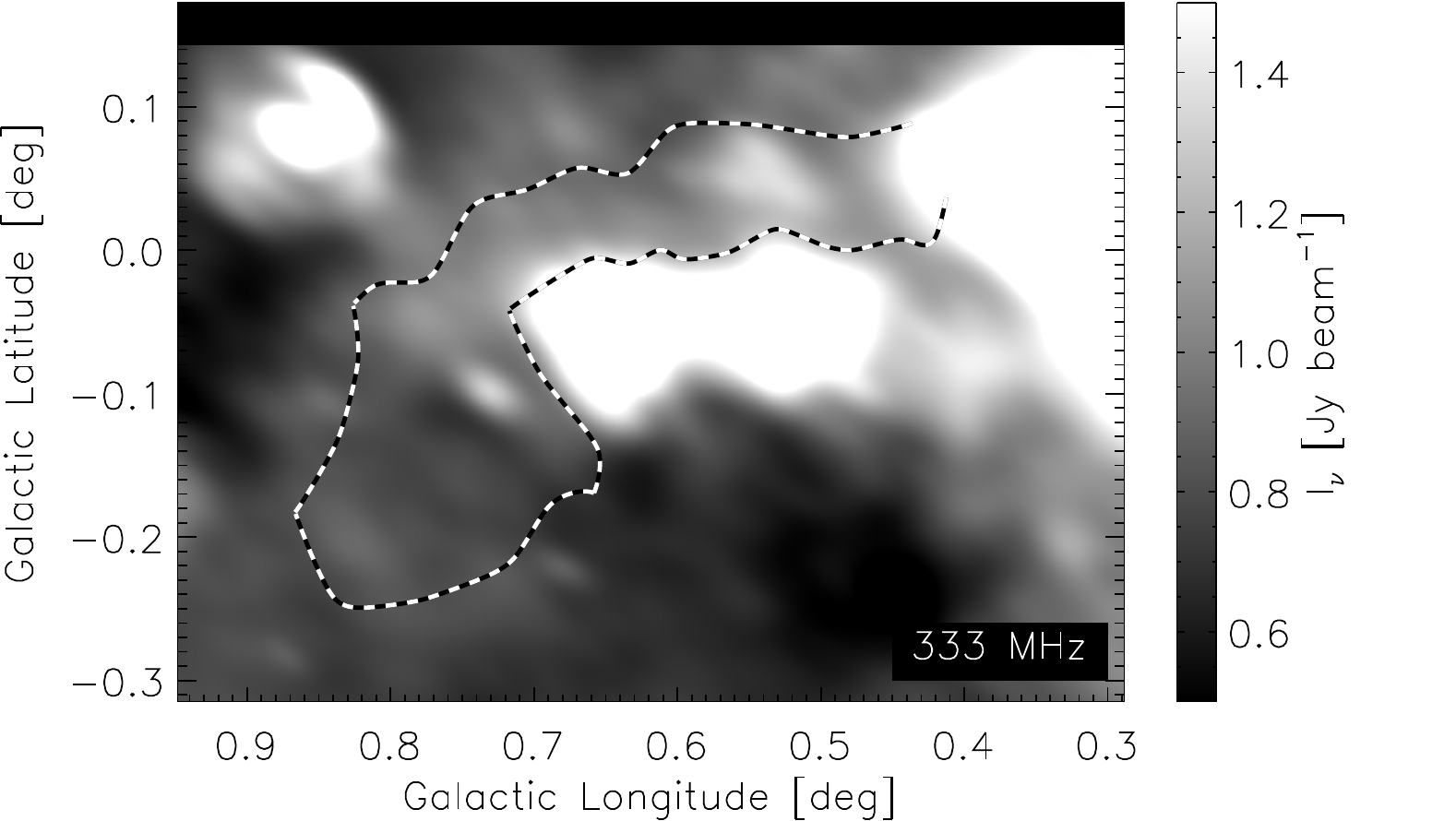}
    \vspace{0.25in} 
\includegraphics[scale=0.3,angle=90]{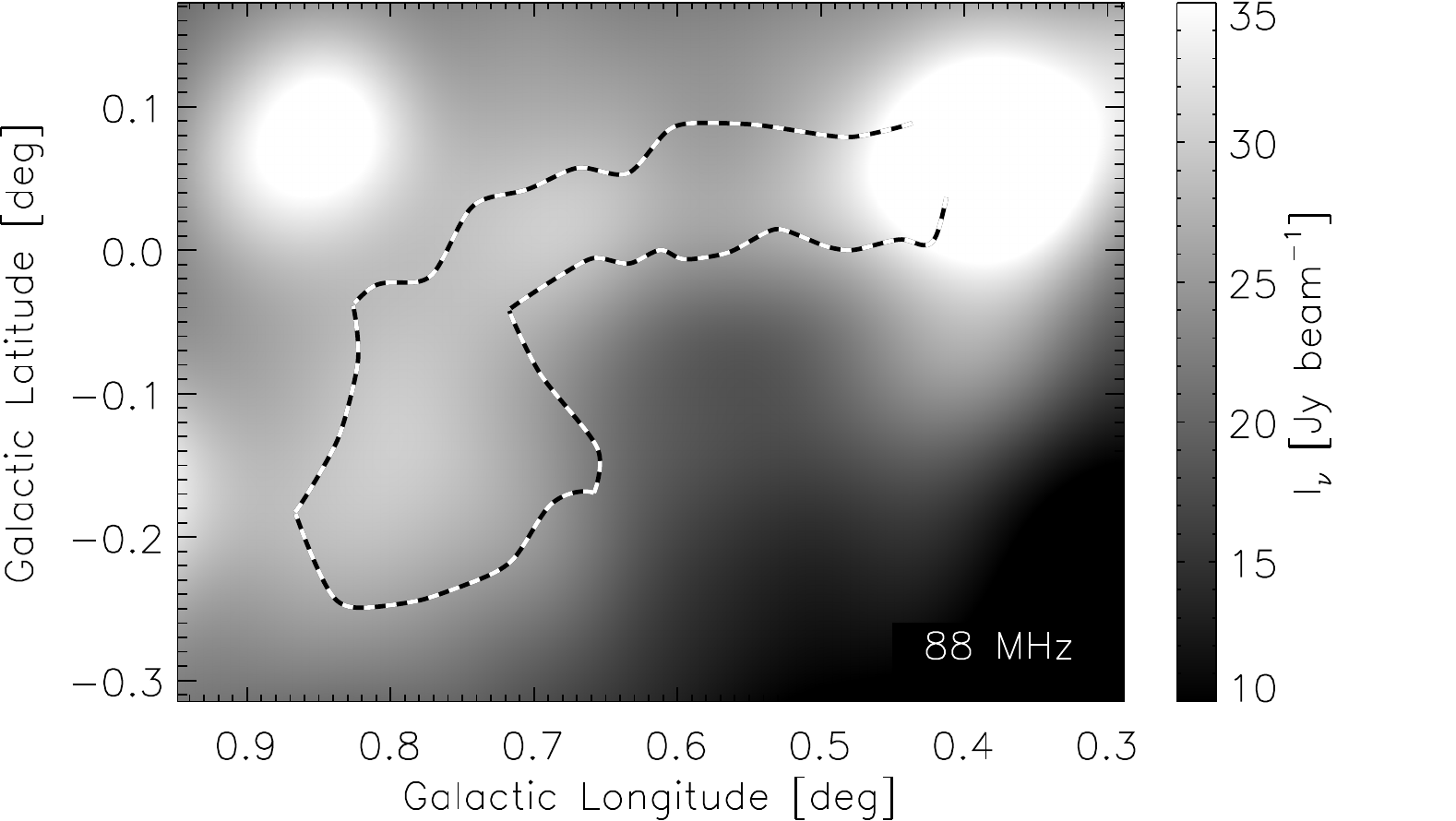}
 \hspace{0.5in} 
   \includegraphics[scale=0.3,angle=90]{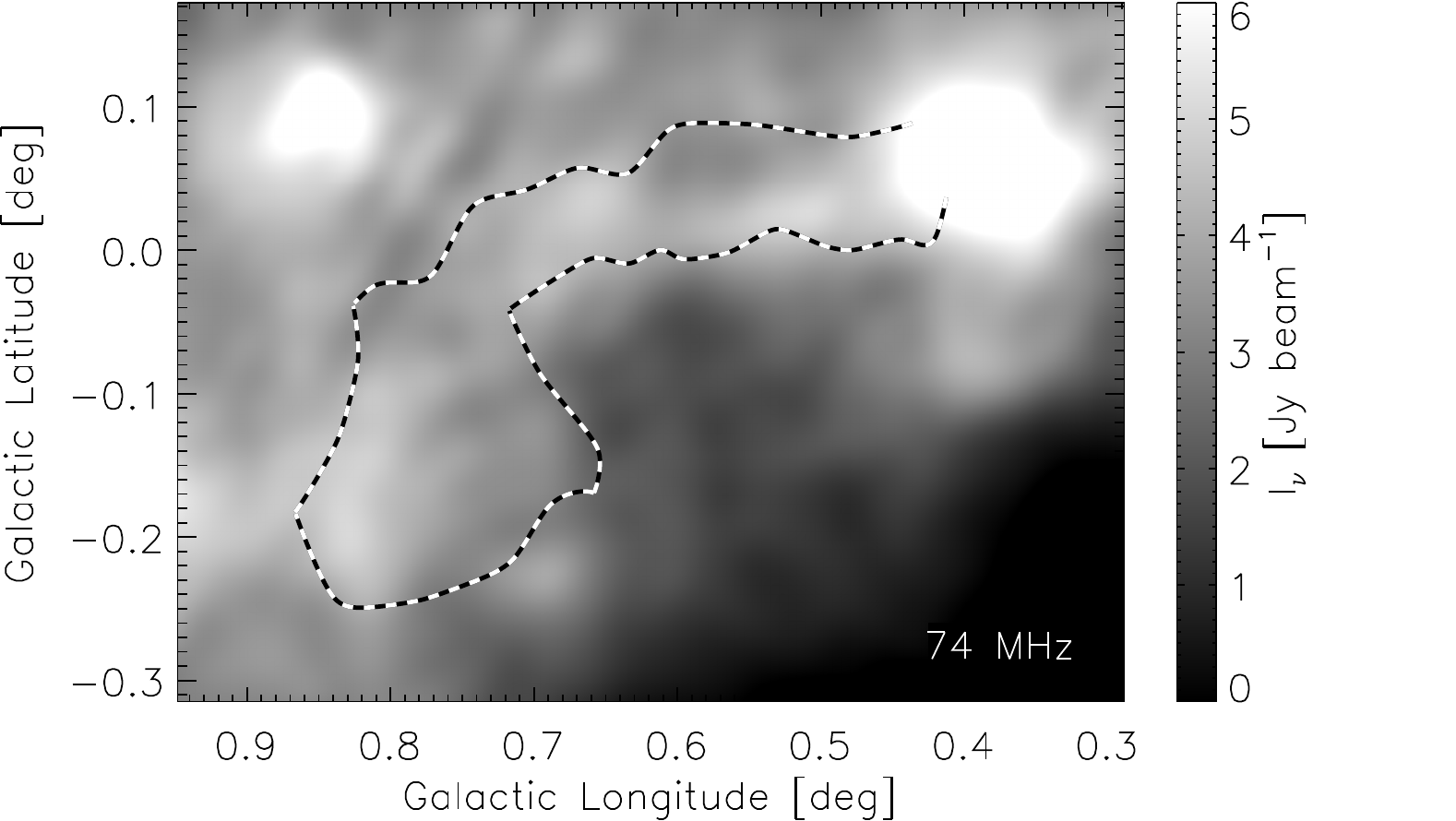}
\caption{
%\small
Images of Sgr B in 6 panels.
The dashed   contour  in all 6 panels outlines
the hook-shaped region to the north of  Sgr B1 and Sgr B2 where
correlation between diffuse 1280 MHz  emission  and 8 $\mu$m extinction, hydrogen column density, and 74 MHz emission  is detected.
The color bar to the right of each image shows the scale for each image. 
{\it  (a)}
An  8$\mu$m {\it Spitzer} image shows the Sgr B IRDC. 
{\it  (b)}
$N_{\rm H}$ column density map derived from far-IR emission of dust from Hershel SPIRE data. 
The black spot is a region of no data due to saturation of the SPIRE 
detectors in one or more bands.
{\it (c)}
A 1280 MHz MeerKAT image  with resolution  $4''\times4''$.
The L-band (856$-$1712~MHz) system was used, with the correlator configured to deliver 4,096 frequency
channels. 
 {\it  (d)}
A 333 MHz  MHz image taken by combining VLA and GBT images   with resolution   
125$''\times125''$ \citep{2006AJ....132..242N,2005ApJ...626L..23L}. 
 {\it  (e)}
An 88  MHz image taken with MWA  with bandwidth  72-103 MHz and resolution  
$327.86''\times309.92''$ (position angle$=-70.19^\circ$ \citep{2022PASA...39...35H}. 
The emission at this frequency is
mainly due to nonthermal radiation  since all thermal features are optically thick and appear in  absorption against 
the background synchrotron  emission.
 {\it  (f)} 
The low-frequency  74 MHz  VLA image with a resolution of 
125$''\times125''$ \citep{2003ANS...324...17B,2005ApJ...626L..23L,2006AJ....132..242N}.
74 MHz VLA data. 
The morphology  is  similar to the 88 MHz image with  thermal features such as Sgr B appearing in absorption. 
}
\end{figure}

%\begin{figure}[ht!]
\begin{figure}
%\begin{figure}[p]
%\centering
%\epsscale{0.2}
%\hspace{1.0in}
\centering
   \includegraphics[scale=0.3,angle=90]{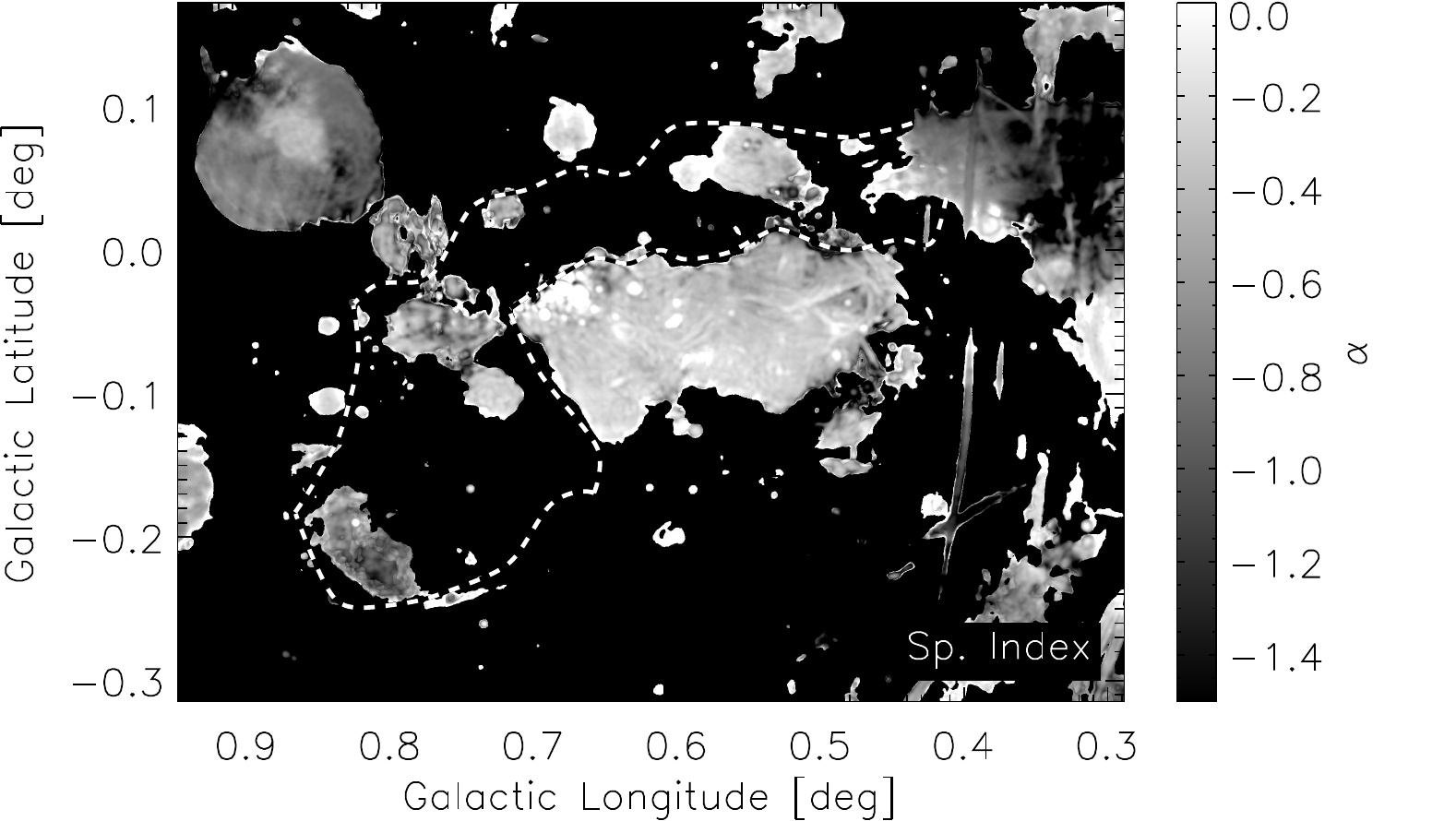}
    \vspace{0.25in} 
   \includegraphics[scale=0.3,angle=90]{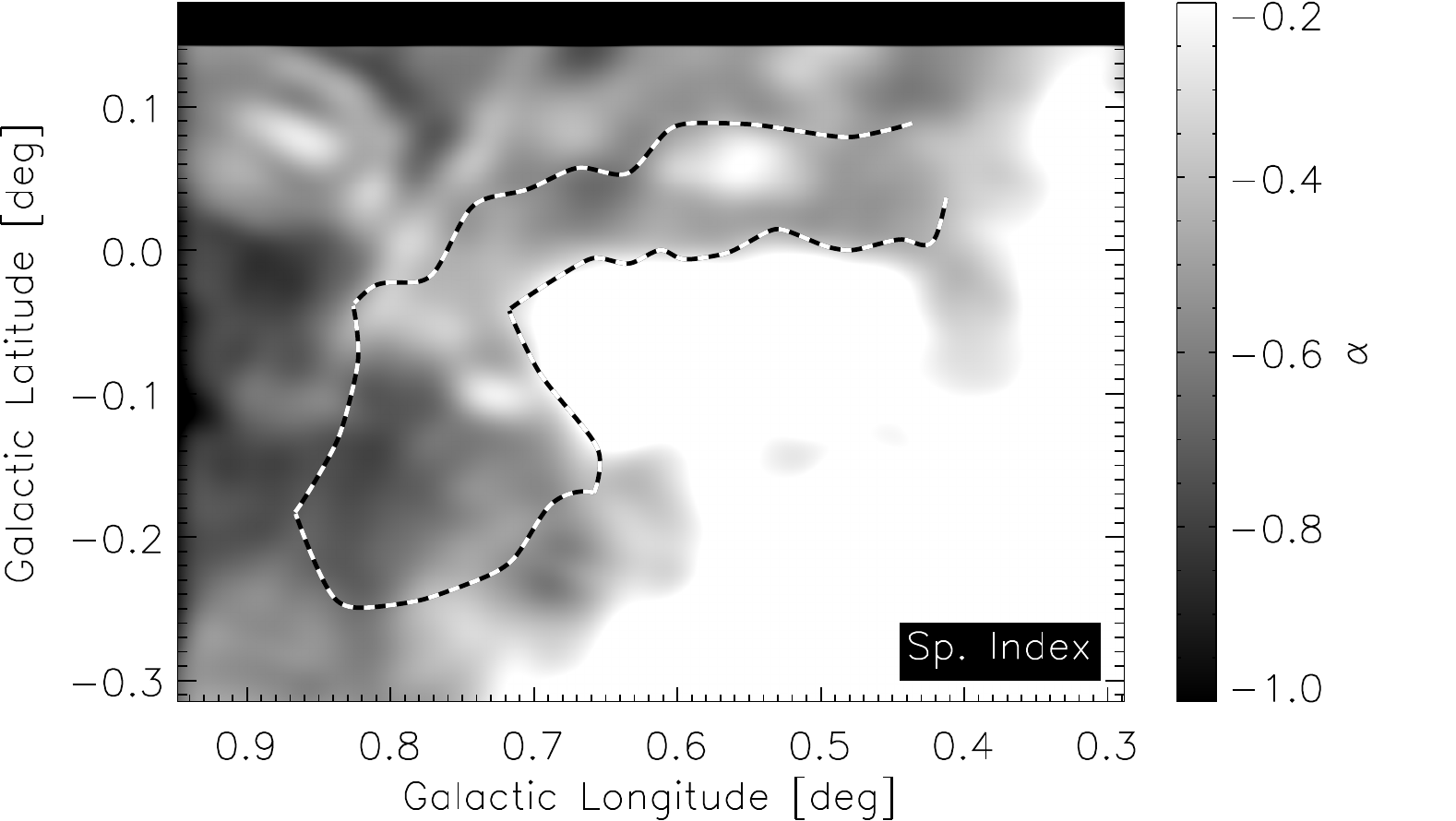}
    \vspace{0.25in} 
   \includegraphics[scale=0.3,angle=90]{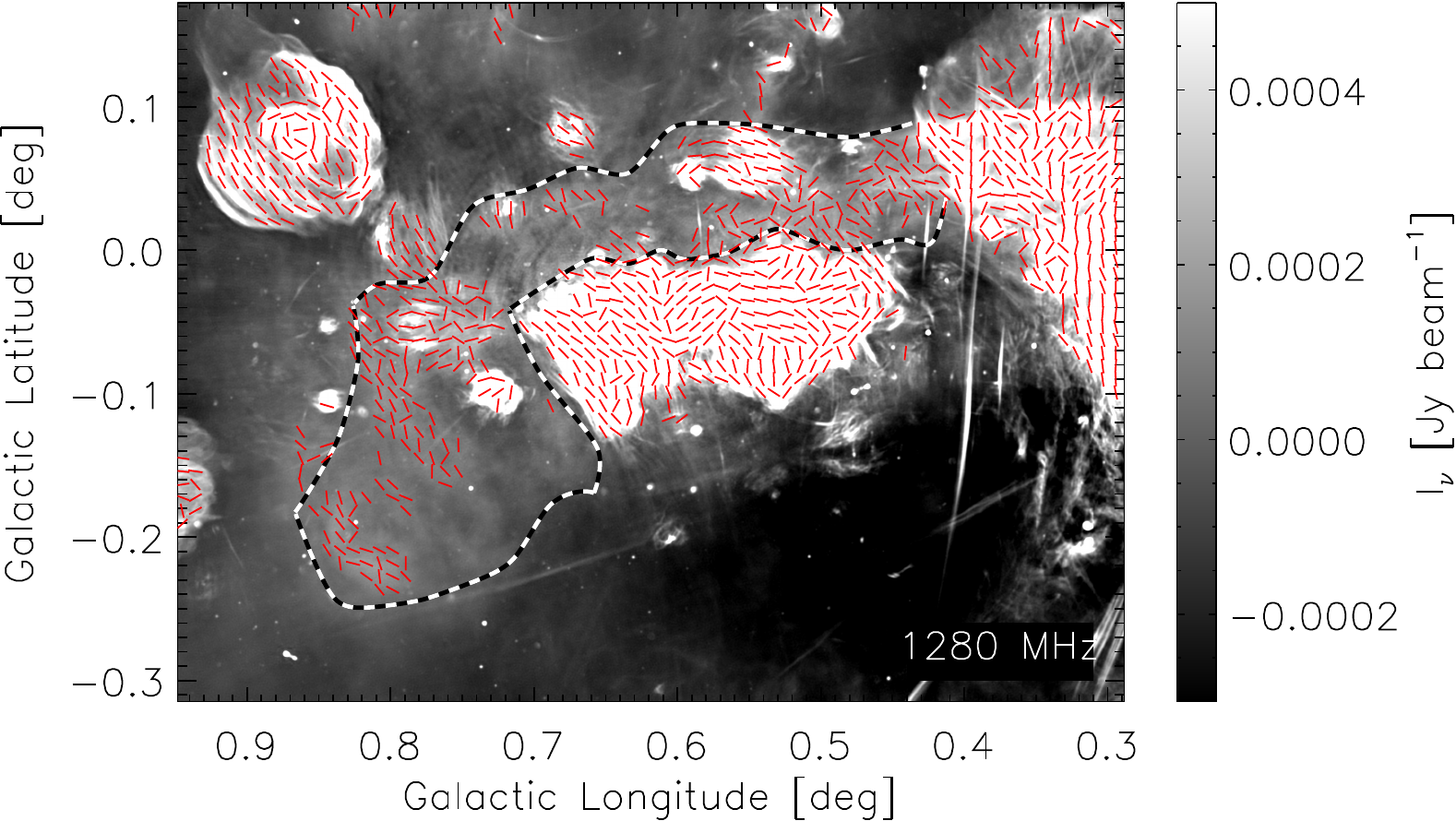}
\caption{
\small
{\it (a)}
The spectral index image  of  Sgr B was
derived using 16 $\times$53.5 MHz  frequency sub-bands   spread between  the lowest and highest frequency sub-bands 0.882 and 1.685 GHz
and with a resolution of $\sim8''$ \citep{2022ApJ...925..165H}.
The spectral index to  the southeast  of the hook  has  a mean 
$\alpha=-0.9\pm0.2$.
The region to the northwest has mean  $\alpha\sim-0.6\pm0.2$.  
{\it  (b)}
The spectral index between 333 and
 74 MHz is estimated from  VLA images taken with a spatial resolution of $125''\times125''$. 
The mean spectral
index within the contour to the southeast is $\sim-0.7\pm0.1$ where the linear feature to the north of Sgr B 
has a mean spectral index $\sim-0.5\pm0.1$.  
The color bar to the right of each image  shows the scale for each image. 
{\it  (c)}
 Magnetic field orientation calculated using the SIGs technique \citep{2017ApJ...842...30L}.
Bars indicate the local magnetic field direction, but the lengths of the bars are not  indicative of magnetic 
field strength. The size of each  pixel is 1.1$''$ and the resolution of the image is 4$''$ 
with the  image size of  2161$\times$1604 pixels. 
The  SIG-inferred magnetic field image  achieves a resolution of  $\sim22''$. 
}
\end{figure}

%\begin{figure}
\begin{figure}
\centering
%\epsscale{0.1}
   \includegraphics[scale=0.5,angle=0]{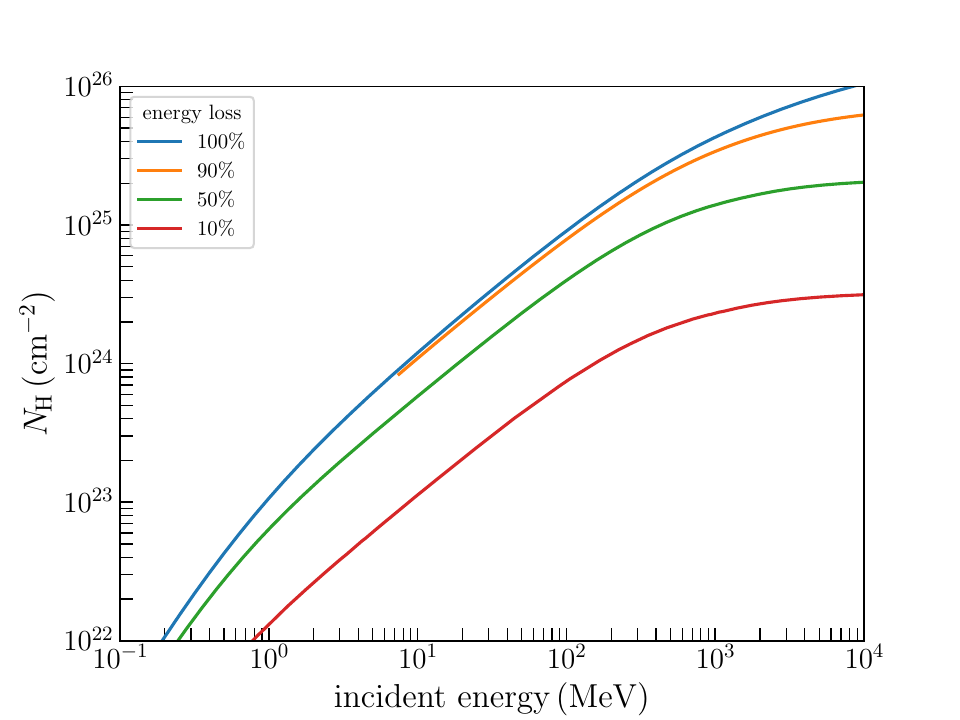}
   \includegraphics[scale=0.5,angle=0]{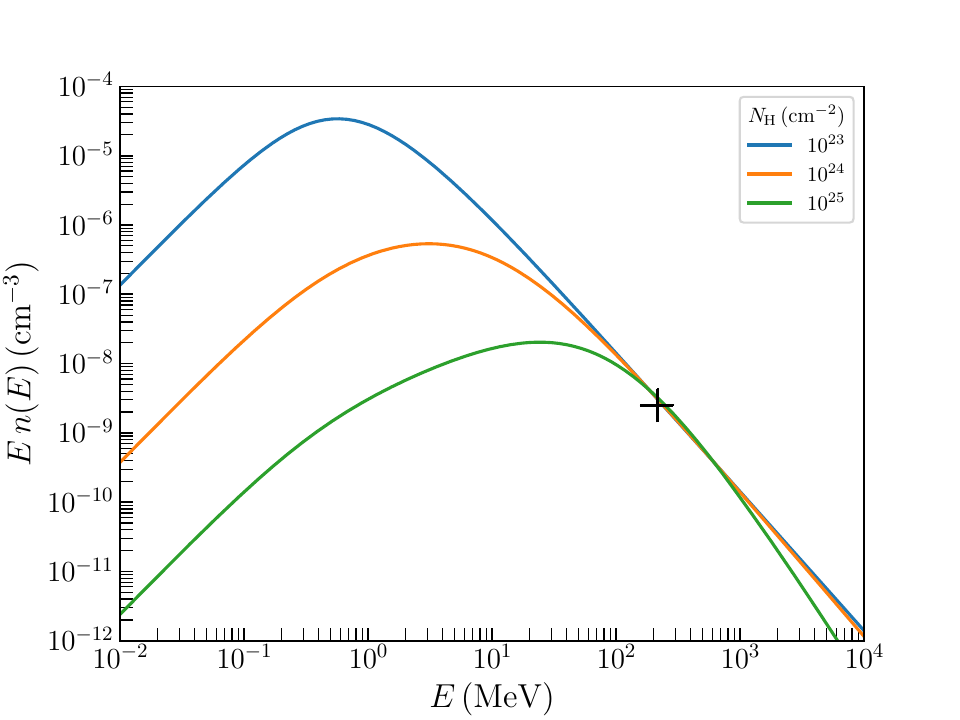}
   \includegraphics[scale=0.5,angle=0]{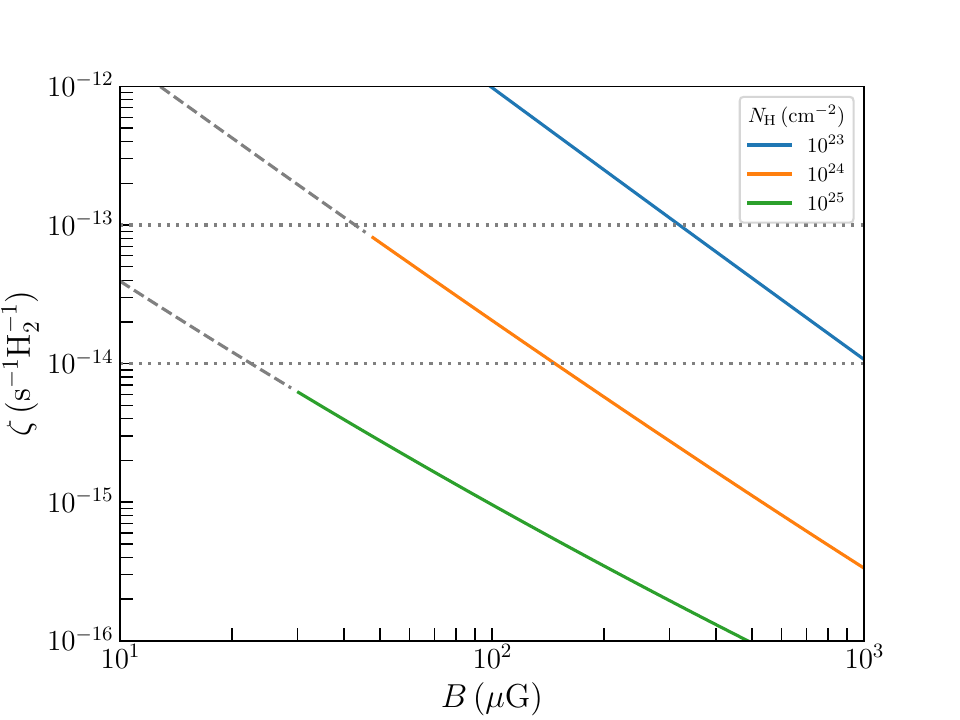}
\caption{
\small
{\it  (a)}
Blue  curve shows the electron range in interstellar gas as a function of incident kinetic energy.  The range is usually expressed in g\,cm$^{-2}$ but 
here has been converted 
to the equivalent hydrogen column (cm$^{-2}$).  The additional curves show the partial range corresponding to a given fractional energy loss.
{\it  (b)}
Computed cosmic-ray electron spectra for $B=100\,\mu$G and different column densities.  The unattenuated energy spectrum is assumed to be a 
power law, with normalization 
and slope adjusted to match the observed synchrotron emissivity and spectral index at 74\,MHz.  The black cross indicates the 
initial estimate of $En(E)$ at the 
characteristic electron energy $E_\nu$ of the electrons that strongly emit at 74\,MHz 
(see eqs \ref{eqn:Enu}--\ref{eqn:jnu}). 
{\it  (c)} 
Cosmic-ray-electron ionization rate per hydrogen molecule implied by the observed synchrotron emissivity of the halo of 
Sgr B2, as a function of the cloud's magnetic field strength and for column densities $N_\mathrm{H}=10^{23-25}$\,cm$^{-2}$.  The curves are dashed at 
low field strengths, 
where the electron energy density exceeds that of the magnetic field.  The horizontal grey dotted lines bracket the typical range of cosmic-ray ionization rates in the 
inner Galaxy (see text). 
}
\end{figure}

\begin{figure}
\centering
%\epsscale{0.1}
   \includegraphics[scale=0.3,angle=90]{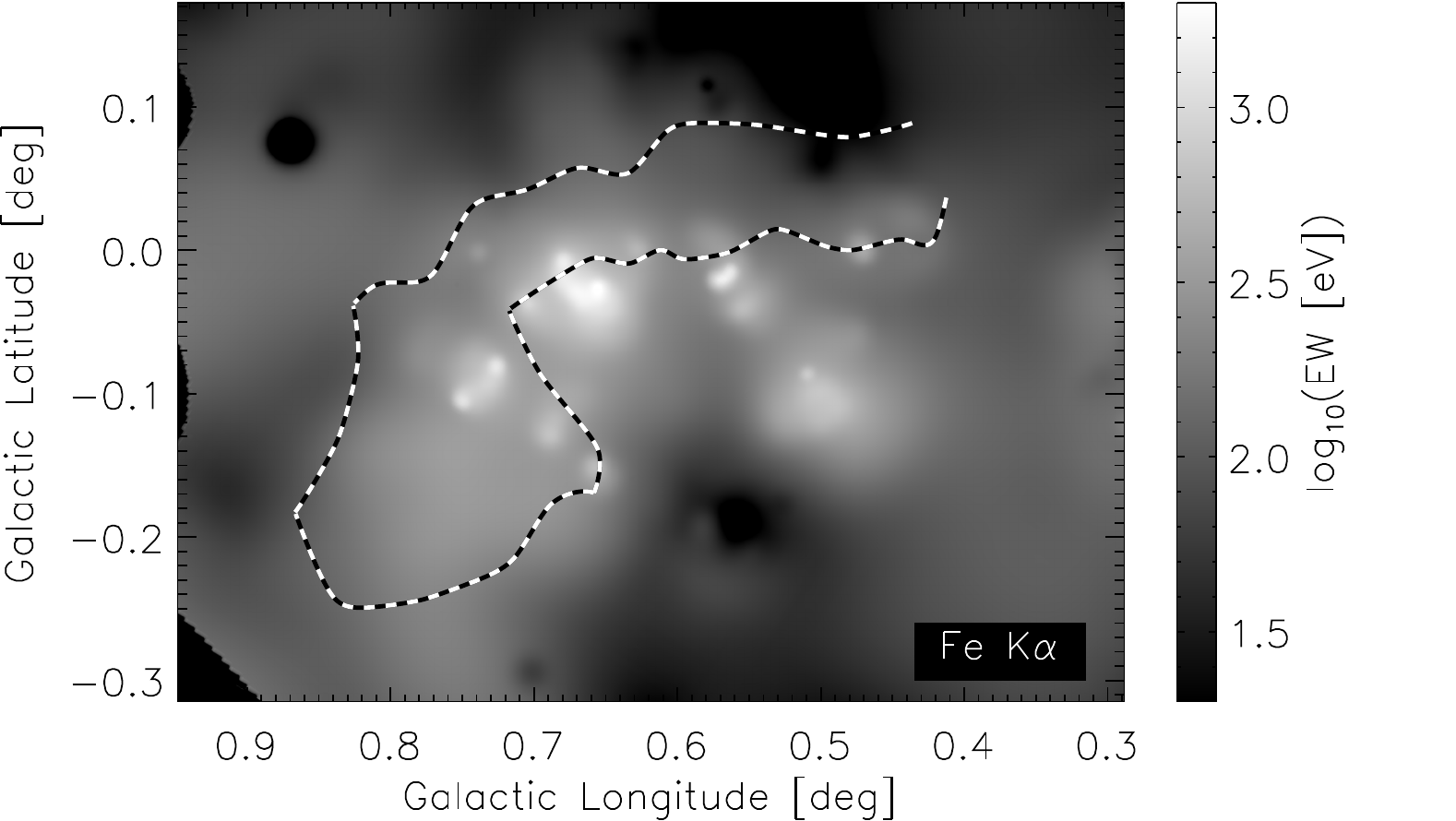}
   \includegraphics[scale=0.3,angle=90]{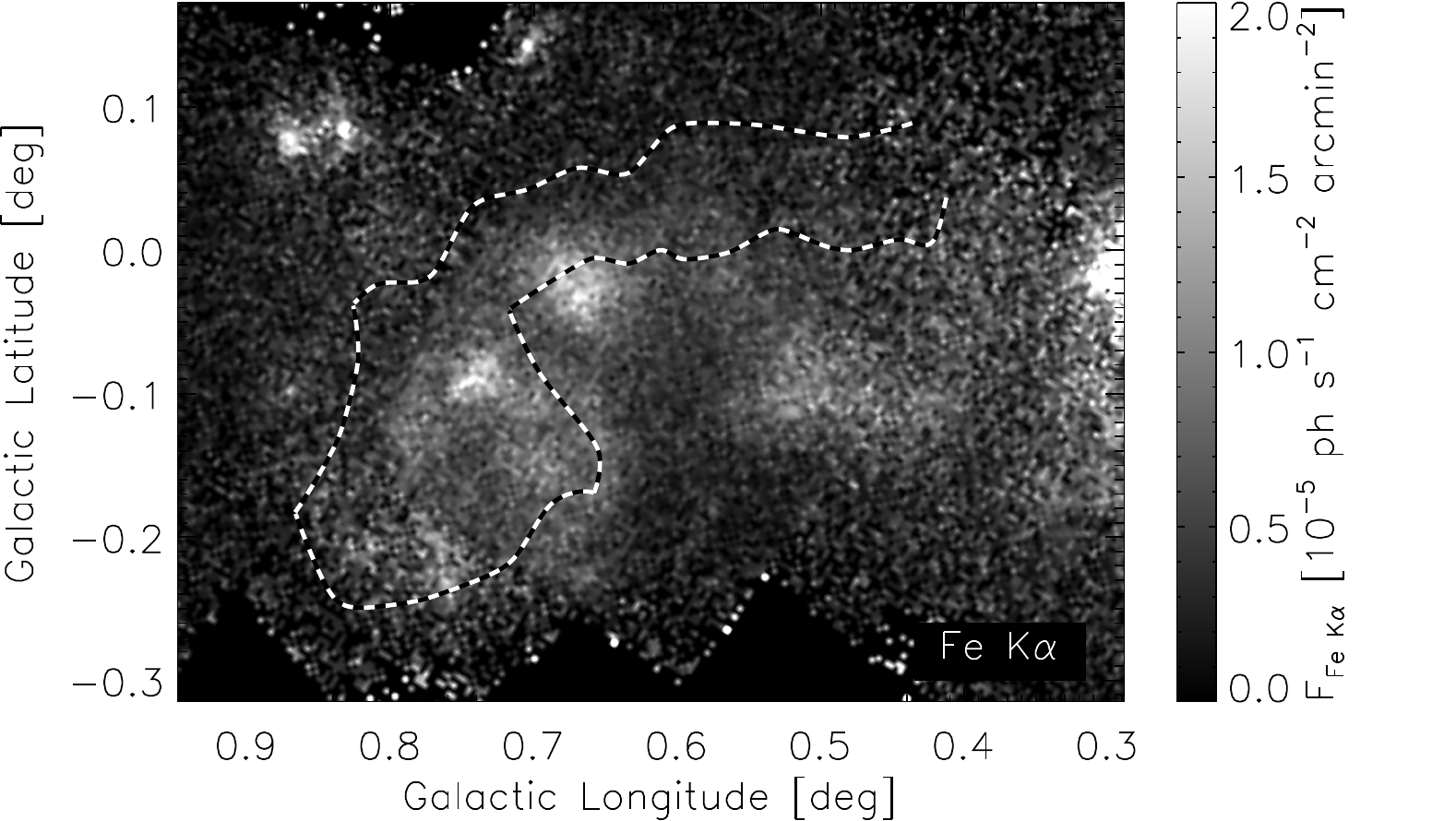}
   \includegraphics[scale=0.5,angle=0]{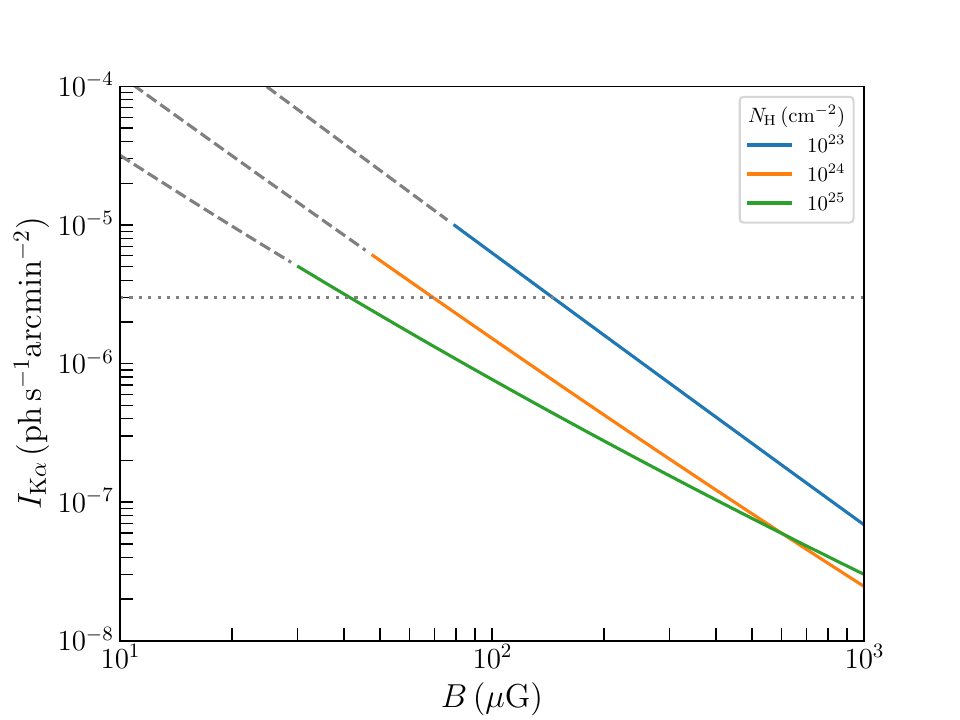}
   \includegraphics[scale=0.5,angle=0]{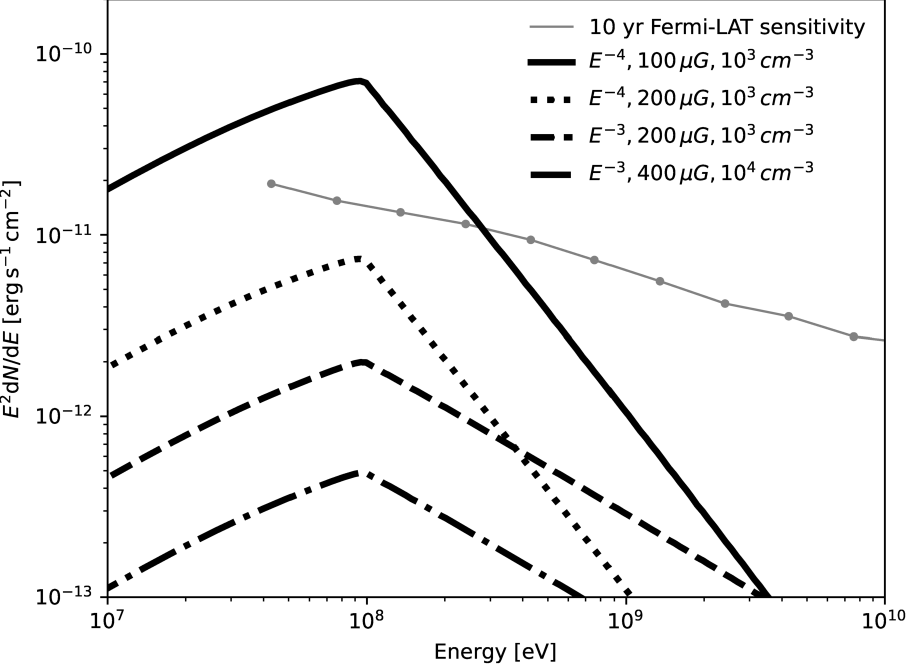}
 \caption{ \small 
{\it (a)} The equivalent width (EW) of the Fe K$\alpha$ line 
emission from the Sgr B2 complex at 6.4 keV. 
Diffuse emission to the SE traces the dark IRDC hook 
\citep{2007ApJ...656..847Y}. 
{\it (b)} 
The intensity of 6.3-6.5 keV line emission based on Suzaku observations 
 \citep{2009PASJ...61S.255K}. 
The color bar to the right of each image shows the scale.
{\it (c)} 
The intensity of the 6.4\,keV iron K$\alpha$ line as a function of the magnetic field  for an iron abundance of twice solar.  
The horizontal grey dotted line indicates the 
typical intensity, as shown in {\it (a)} and {\it (b)}. 
{\it (d)} 
Predicted $\gamma$-ray spectra from the same electron population due to non-thermal bremsstrahlung radiation. It is clear 
that 100$\mu$G magnetic field with energy spectrum $E^{-4}$ is ruled out\citep{2021ApJS..256...12A}. 
} 
\end{figure}

\label{lastpage}

%\begin{thebibliography}
%\expandafter\ifx\csname natexlab\endcsname\relax\def\natexlab#1{#1}\fi
%\end{thebibliography}
\end{document}